\documentclass[prb, reprint]{revtex4-1} 

\usepackage{amsmath}  % needed for \tfrac, \bmatrix, etc.
\usepackage{mathtools} % for partial derivatives
\usepackage{diffcoeff}  % for partial derivatives math
\usepackage{amsfonts} % needed for bold Greek, Fraktur, and blackboard bold
\usepackage{graphicx} % needed for figures
\usepackage{float}
\usepackage{color}
\usepackage{ragged2e} % for justifying captions
\usepackage[justification=justified, format=plain]{caption}
\usepackage[justification=justified, format=plain]{subcaption}
\usepackage{multirow}

\usepackage{lineno, blindtext}
\usepackage{xr-hyper}
\begin{document}

\title{Measurement of electromagnetic radiation force using a capacitance-bridge interferometer}

\author{Devashish Shah}
\affiliation{Department of Physics, Indian Institute of Technology Bombay, Mumbai, India}

\author{Pradum Kumar}
\affiliation{Department of Physics, Indian Institute of Technology Bombay, Mumbai, India}

\author{Pradeep Sarin}
\affiliation{Department of Physics, Indian Institute of Technology Bombay, Mumbai, India}
\date{\today}

%-----------------------------------------------------------------------%

\begin{abstract}
We present a mechanical cantilever-based tabletop interferometer to measure the radiation force exerted by light. Using a high-power ($\sim 1\,\mathrm{W}$) pulsed laser beam, we excite mechanical oscillations in a thin metallic cantilever. The cantilever forms a parallel-plate capacitor with a printed circuit board trace. Using a capacitance-bridge geometry, we measure small capacitance changes of the order of femto-farads, induced by the radiation forces of a few nano-newtons. This experiment uses equipment commonly found in an undergraduate teaching laboratory for physics and electronics while providing insight into electromagnetic wave theory, circuit design for low-noise measurements, and Fourier analysis.
\end{abstract}

\date{\today}
\maketitle

%\begin{linenumbers}
\section{Introduction}
Students typically learn about electromagnetic radiation, radiation momentum-density, and radiation force in advanced courses on electromagnetism. These concepts often remain purely theoretical, with some reason: the unambiguous quantification of radiation forces acting on macroscopic objects has historically been difficult. \cite{brush1969radiometer,lebedev1883experimental} 
In recent years, however, measurements of these forces have been achieved using micromachined resonators \cite{dakang, Boales2017, russell_measuring_2025} and precision optics.\cite{Partanen2021}
In this paper, we present a novel experiment to determine the radiation force exerted by a pulsed laser source on a thin metal strip suspended over a printed circuit board (PCB) trace in ambient air. 
The experiment can be performed using equipment typically found in an undergraduate teaching laboratory. 

The momentum density of an electromagnetic (EM) plane wave traveling in vacuum with electric field amplitude $E_0$ is:
\begin{equation}
    \vec{g}_{\text{Rad}} =\frac{\epsilon_0}{c}E_0^2 \hat{z} 
    \label{eq: Radiation momentum density},
\end{equation}
$c$ being the speed of light and $\epsilon_0$ the permittivity in free space, with $\hat{z}$ being the direction of propagation. Consequently, for a plane EM wave incident normally on a surface of area $\mathcal{A}$, the radiation force is: 
\begin{equation}
    {F}_{\text{Rad}}=\frac{I\mathcal{A}}{c}(2r(\lambda )+a(\lambda)).
    \label{eq: radn pressure}
\end{equation}
Here, $I=c\epsilon_0 E_0^2/2$ is the time-averaged intensity. The incident light is partly absorbed and reflected with associated absorption and reflection coefficients, $a(\lambda)$ and $r(\lambda)$, which are wavelength-dependent.

For a laser power of $1\,\text{W}$, we expect the radiation force to be a few $\mathrm{nN}$. In this paper, we describe an experimental setup that allows to measure these small forces. Section II describes the circuit theory and cantilever dynamics forming the basis of the experiment. Section III describes the experimental setup, followed by results and analysis in Section IV. Analytical calculations for the flexure of a cantilever are provided in Appendix A, followed by the description of electrical layout and design considerations in Appendix B.
%-----------------------------------------------------------------------%
\section{Circuit Theory AND CANTILEVER DYNAMICS}
\begin{figure}[ht]
\nolinenumbers
  \centering
  \begin{subfigure}[b]{0.9\linewidth}
      \includegraphics[width=\linewidth]{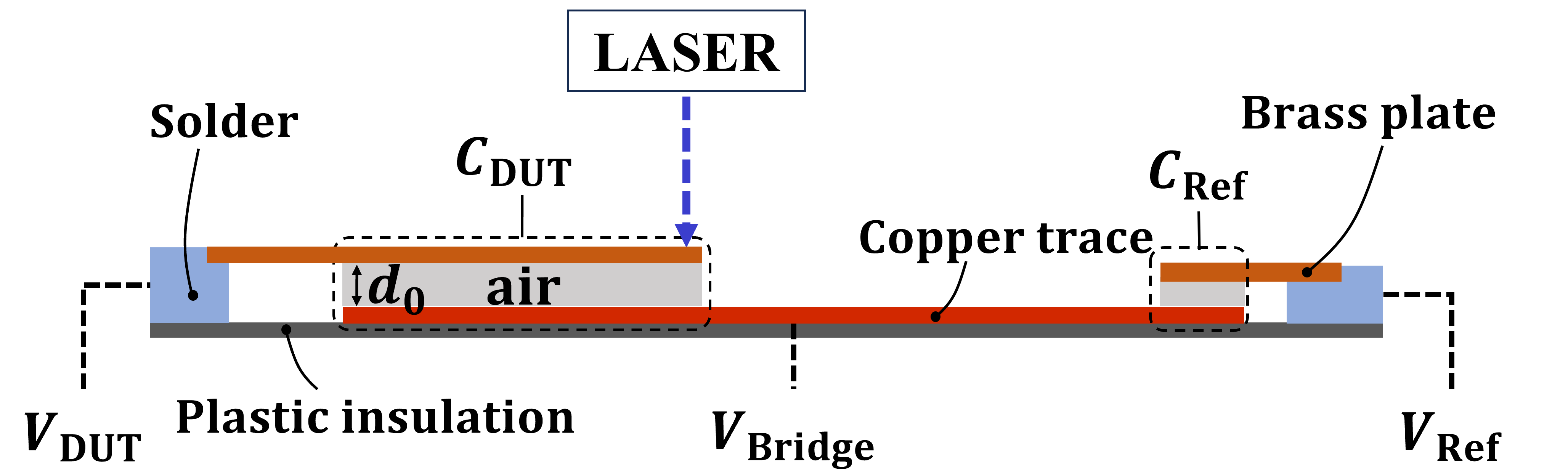}
      \caption{}
    \end{subfigure}
  \begin{subfigure}[b]{0.8\linewidth}
      \includegraphics[width=\linewidth]{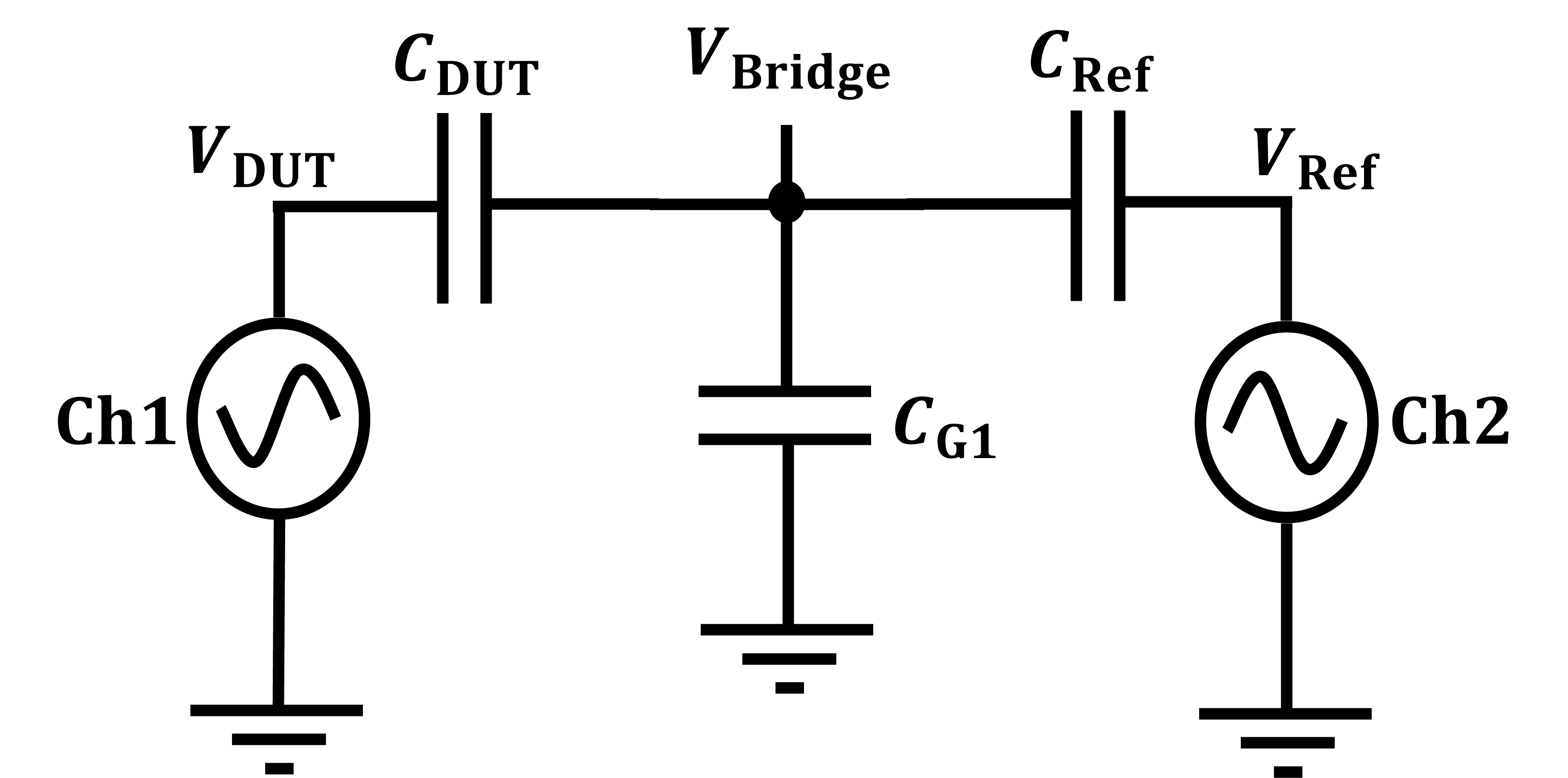}
      \caption{}
    \end{subfigure}
    \caption{\justifying{
        (a) Physical layout of the capacitance bridge consisting of two air capacitors, made with brass cantilevers parallel to the PCB traces with plate separations less than $1\,\mathrm{mm}$. (b) Schematic of the capacitance bridge circuit. The voltage amplitudes of the signal are used to label the nodes in the circuit.}}
    \label{fig:bridge schematics}
\end{figure}

The device under test (DUT) is a thin brass strip in parallel to a copper PCB trace, forming an air capacitor $C_{\text{DUT}}$ (Fig. \ref{fig:bridge schematics}a). One end of the strip is soldered to the input signal terminal connected to channel-1 (Ch1 in Fig. \ref{fig:bridge schematics}b) of an arbitrary function generator (AFG). A reference air capacitor $C_{\text{Ref}}$, formed using a similar metal strip in parallel with the PCB trace, is soldered to the input signal terminal connected to channel-2 (Ch2 in Fig. \ref{fig:bridge schematics}b) of the AFG. The capacitors $C_{\text{DUT}}$ and $C_{\text{Ref}}$ are mounted in a ``capacitance bridge" configuration (Fig. \ref{fig:bridge schematics}a). The metal strip for $C_{\text{Ref}}$ is kept short and broad to keep it rigid and fixed. $C_{\text{G1}}$ is the lumped parasitic capacitance to ground at the bridge junction labeled $V_{\text{Bridge}}$ in Figure.\ref{fig:bridge schematics}b. \footnote{We work with sinusoidal signals at a fixed carrier frequency signal $f_C$. Thus, we express the voltage at any circuit terminal as $V = V_\text{N}\sin(2\pi f_Ct + \delta_\text{N})$, $V_\text{N}$ being the amplitude modulating the oscillations at $f_C$, with a phase shift of $\delta_\text{N}$. This voltage amplitude is used to label each node, and is the quantity of interest for the experiment.}
\begin{figure}[htbp]
\nolinenumbers
\centering
      \includegraphics[width=0.7\linewidth]{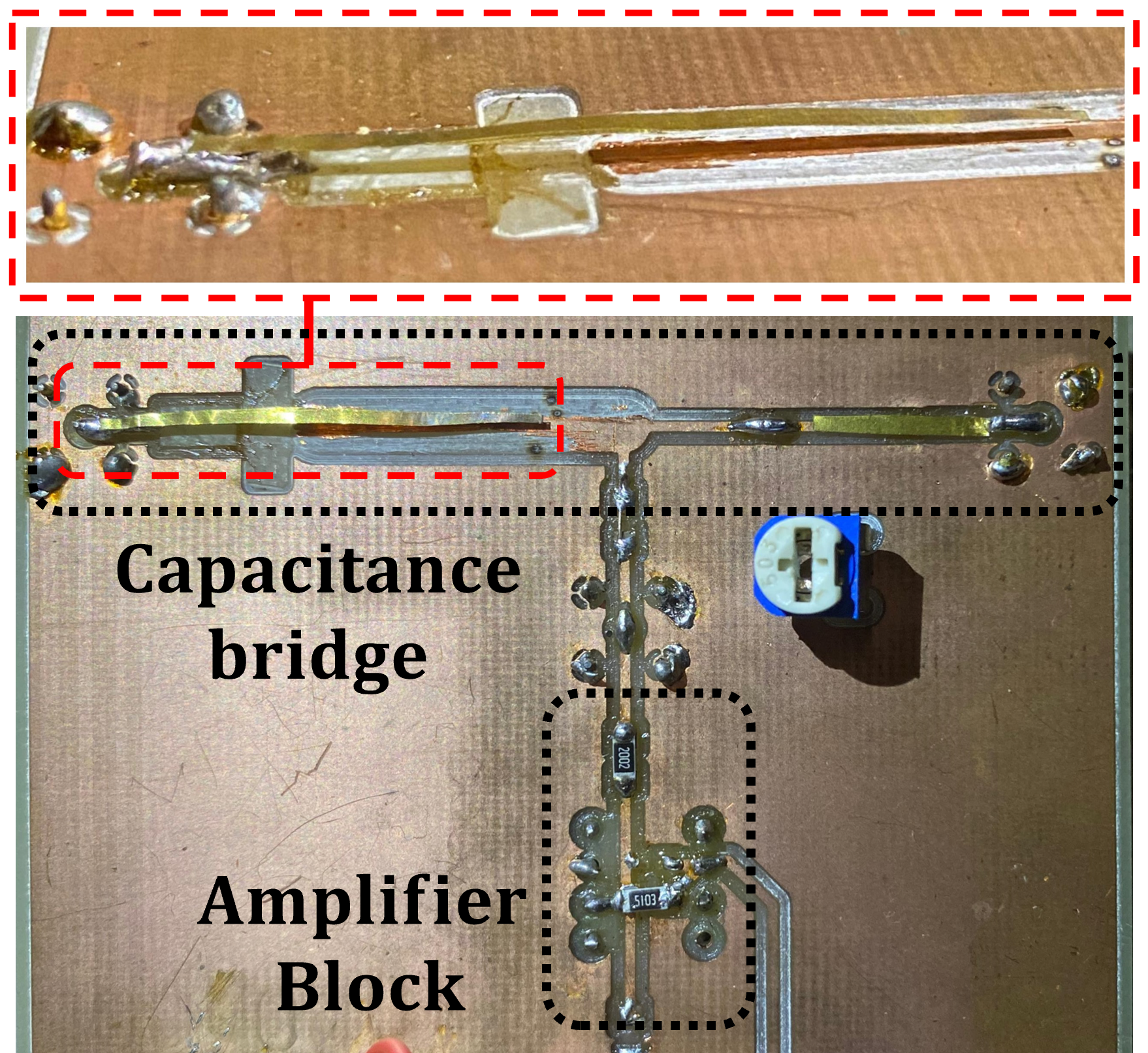}
\caption{\justifying{Image of the assembled PCB, with the capacitance bridge circuit and an integrated Op-Amp (LF411C) based inverting amplifier block. Inset: zoomed-in image of $C_{\text{DUT}}$.}}
\label{fig:PCB and Cantilever photos}
\end{figure}

The signals from Ch1 and Ch2 (Fig. \ref{fig:bridge schematics}b) are sinusoidal voltages $V_{\text{DUT}}\sin(\omega_C t)$ and $V_{\text{Ref}}\sin(\omega_C t +\eta)$ with adjustable amplitudes ($V_{\text{DUT}}$ and $V_{\text{Ref}}$), carrier frequency ($f_C=\omega_C/2\pi$), and relative phase ($\eta$), generated using a two-channel Arbitrary Function Generator (AFG). In principle, a single true floating differential sinusoidal voltage can be used if $C_{\text{Ref}}=C_{\text{DUT}}$. For the measurement, the phase was set on the AFG such that $\eta = \pi$. The sum of the destructively interfering sinusoidal signals is seen at the junction $V_{\text{Bridge}}$ in Figure. \ref{fig:bridge schematics}b.

The net current into the junction $V_\text{Bridge}$ must be zero, and therefore the charges across the capacitors must add to zero:
\begin{equation}
\begin{split}
(V_\text{Bridge} - V_\text{DUT})C_\text{DUT} + (V_\text{Bridge}+V_\text{Ref})C_\text{Ref}\\ + V_\text{Bridge} C_\text{G1} &= 0,
    \label{eq: Kirchhoff_equation}
\end{split}
\end{equation}
where the common sinusoidal $\sin(\omega_C t)$ dependence cancels and the sign of $V_\text{Ref}$ takes into account the relative phase of $\pi$. Rearranging terms in Eq. (\ref{eq: Kirchhoff_equation}) gives:
\begin{equation}
    C_{T}V_{\text{Bridge}}=C_{\text{DUT}}V_{\text{DUT}}-C_{\text{Ref}}V_{\text{Ref}}
    \label{eq: CV relation}
\end{equation}
where, $C_T =C_{\text{DUT}} + C_{\text{Ref}}+C_{\text{G1}}$ is the combined capacitance. The bridge is balanced when $V_{\text{Bridge}}=0$, thus:
\begin{equation}
    V_{\text{DUT}} = V_{\text{Ref}}\frac{C_{\text{Ref}}}{C_{\text{DUT}}}.
    \label{eq: balance}
\end{equation}
Any change in the DUT capacitance $\Delta C_{\text{DUT}}$ due to external drive then leads to a proportional change in bridge voltage amplitude $\Delta V_{\text{Bridge}}$, with $V_{\text{Ref}}, V_{\text{DUT}},$ and $C_{\text{Ref}}$ held constant. This can be seen as the differential of Eq. (\ref{eq: CV relation}):
\begin{equation}
    \Delta V_{\text{Bridge}} = \frac{\Delta C_{\text{DUT}}}{C_{T}}V_{\text{DUT}}.
    \label{eq: delta cdut at balance}
\end{equation}
When periodically hit by a focused laser pulse, the cantilever oscillates, thus changing $d_0$, the air gap (Fig. \ref{fig:bridge schematics}a), and hence $C_{\text{DUT}}$. $\Delta C_{\text{DUT}}$ is large when the pulsing of the laser is resonant with the cantilever's resonance frequency. The change in the capacitance is used to calculate the force exerted by light on the cantilever.

Analogous to any linear damped driven oscillator, the deflection of the cantilever in the presence of an external drive can be described as a superposition of eigenmodes at resonance frequencies $f_n = \omega_n/2\pi$ (Eq. (\ref{Eq: eigenmodes})). A theoretical estimate of the first harmonic frequency of the flexure of an ideal cantilever can be made following the analysis shown in Appendix \ref{appendix:A Dynamics of a damped driven cantilever}. For the brass cantilever used, the expected first harmonic frequency is:
\begin{equation}
\begin{split}
    f_1^{\text{theory}} & = \frac{1}{2\pi} \omega_1 = \frac{1.875^2}{2\pi}\cdot\left(\frac{YM}{\rho \mathcal{A}_{\text{cs}} L_C^4}\right)^{\frac{1}{2}}\\
    & = 40 \pm 4.9 \,\mathrm{Hz}.
\label{eq: omega_1}
\end{split}
\end{equation}
Here, the brass cantilever has Young's modulus\cite{brass_young} $Y = (1.0\pm0.2)\cdot10^{11}\,\mathrm{N/m^2}$, density $\rho = 7575\pm 19.9\,\mathrm{kg/m^3}$, length $L_C = 2.740\pm 0.002\, \mathrm{cm}$, width $b = 1.00 \pm 0.02\,\mathrm{mm}$, and thickness $h = 50\pm1{\,\mu \mathrm{m}}$.  $M = b^3h/12$ is the area moment of inertia, and $\mathcal{A}_{\text{cs}} = bh$ the cross-sectional area of the cantilever. Here, we account for a large error in $Y$, which depends on the exact proportion of metals in the alloy, and dominates the error in the theoretical estimate of $f_1$. The density was measured for a larger square cutout form the same brass shim stock used to make the cantilevers, with a balance having $100\,\mathrm{\mu g}$ resolution. Dimensions $L_C$ and $b$ of the cantilever (DUT) were measured using a Vernier caliper accurate up to $0.02\,\mathrm{mm}$.
%-----------------------------------------------------------------------%
%---------------------------------------%
\section{Experimental setup}
\subsection{PCB and Laser setup}
\label{subsec: laser and PCB}
\begin{figure*}[ht]
\nolinenumbers
  \centering
      \begin{tabular}[c]{cc}
    \multirow{2}{*}[60pt]{
    \begin{subfigure}{0.385\linewidth}
      \includegraphics[width=\linewidth]{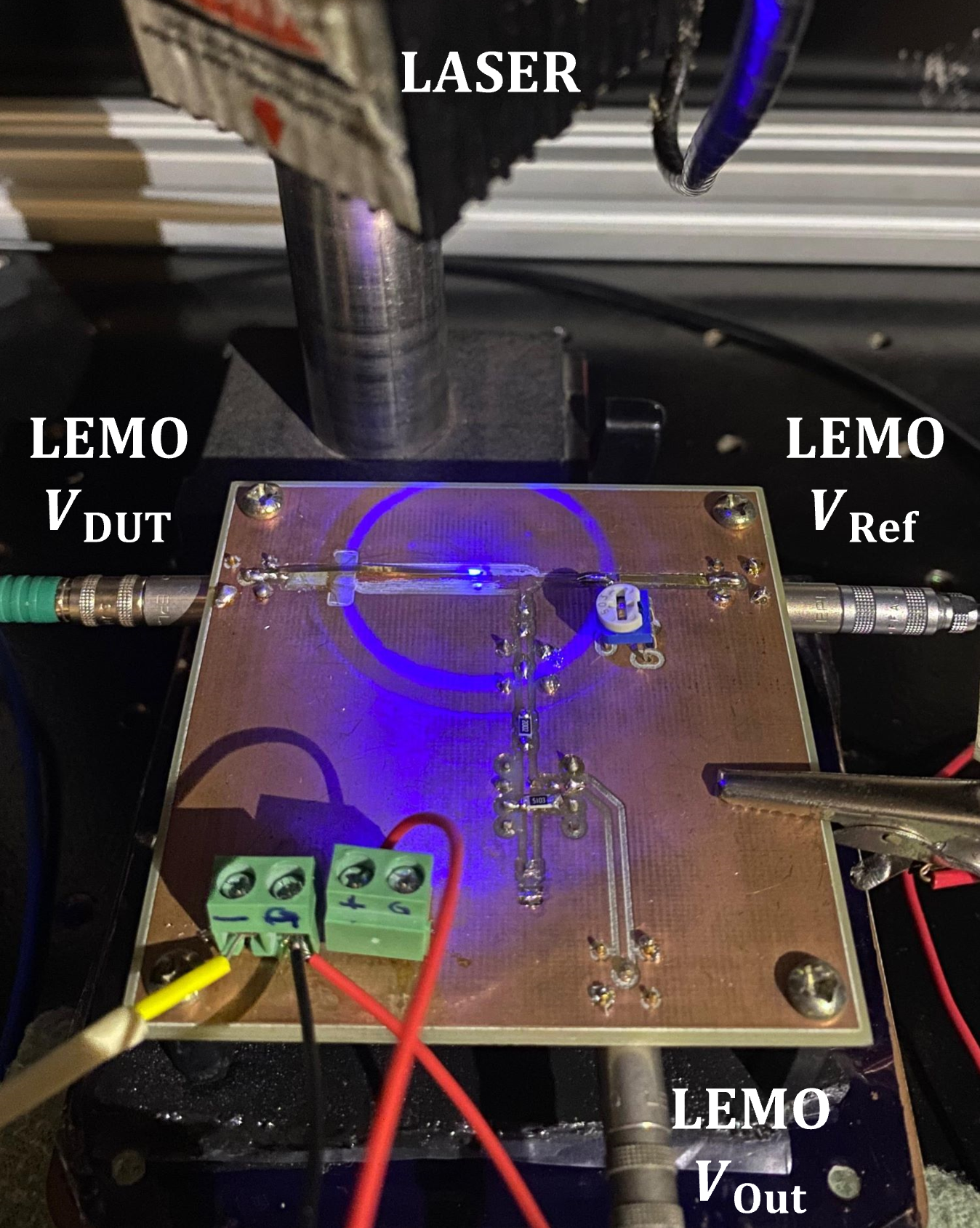}
      \caption{}
    \end{subfigure}}&
   \begin{subfigure}[c]{0.35\linewidth}
      \includegraphics[width=\linewidth]{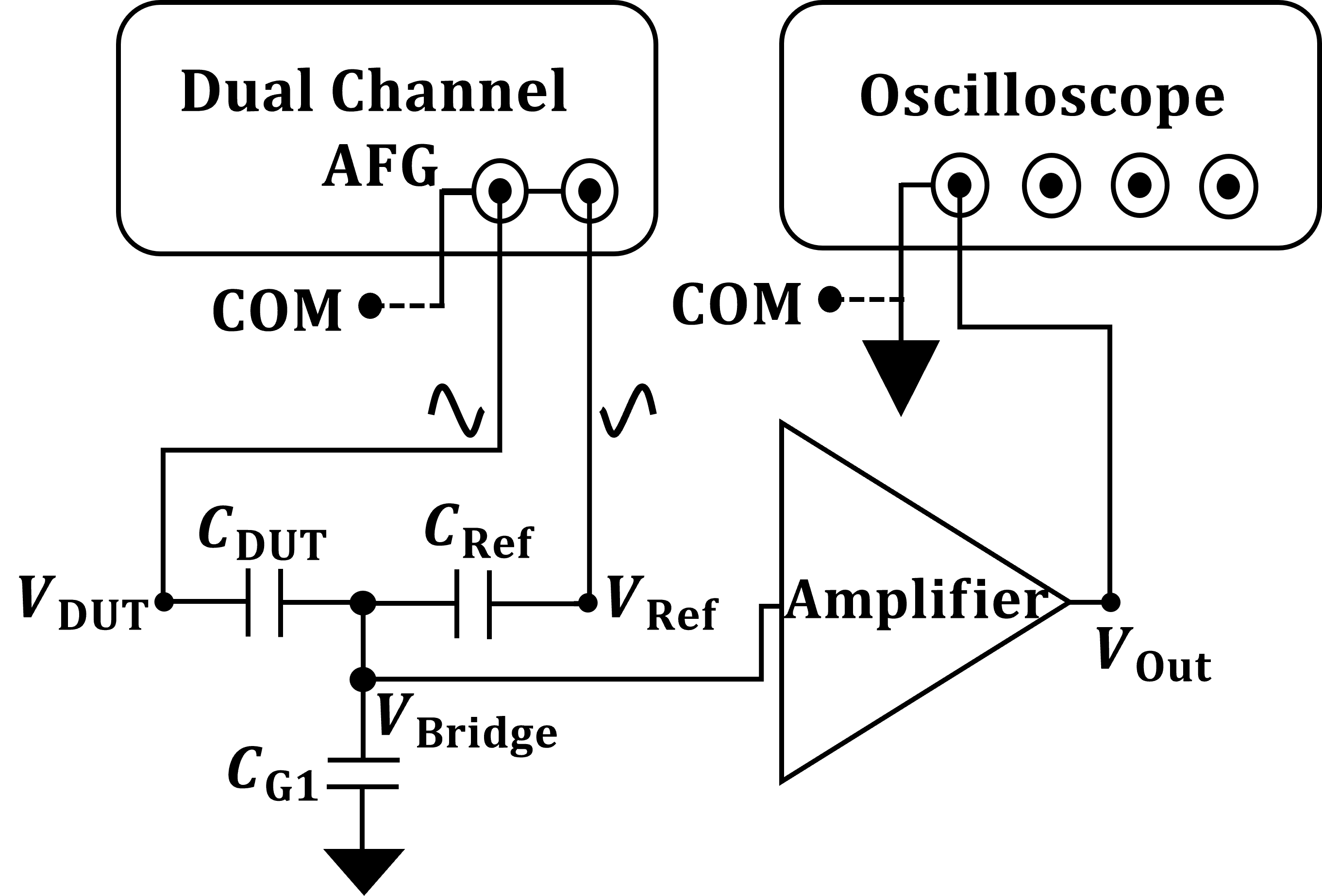}
      \caption{}
    \end{subfigure}\\
    &
    \begin{subfigure}[c]{0.29\linewidth}
      \includegraphics[width=\linewidth]{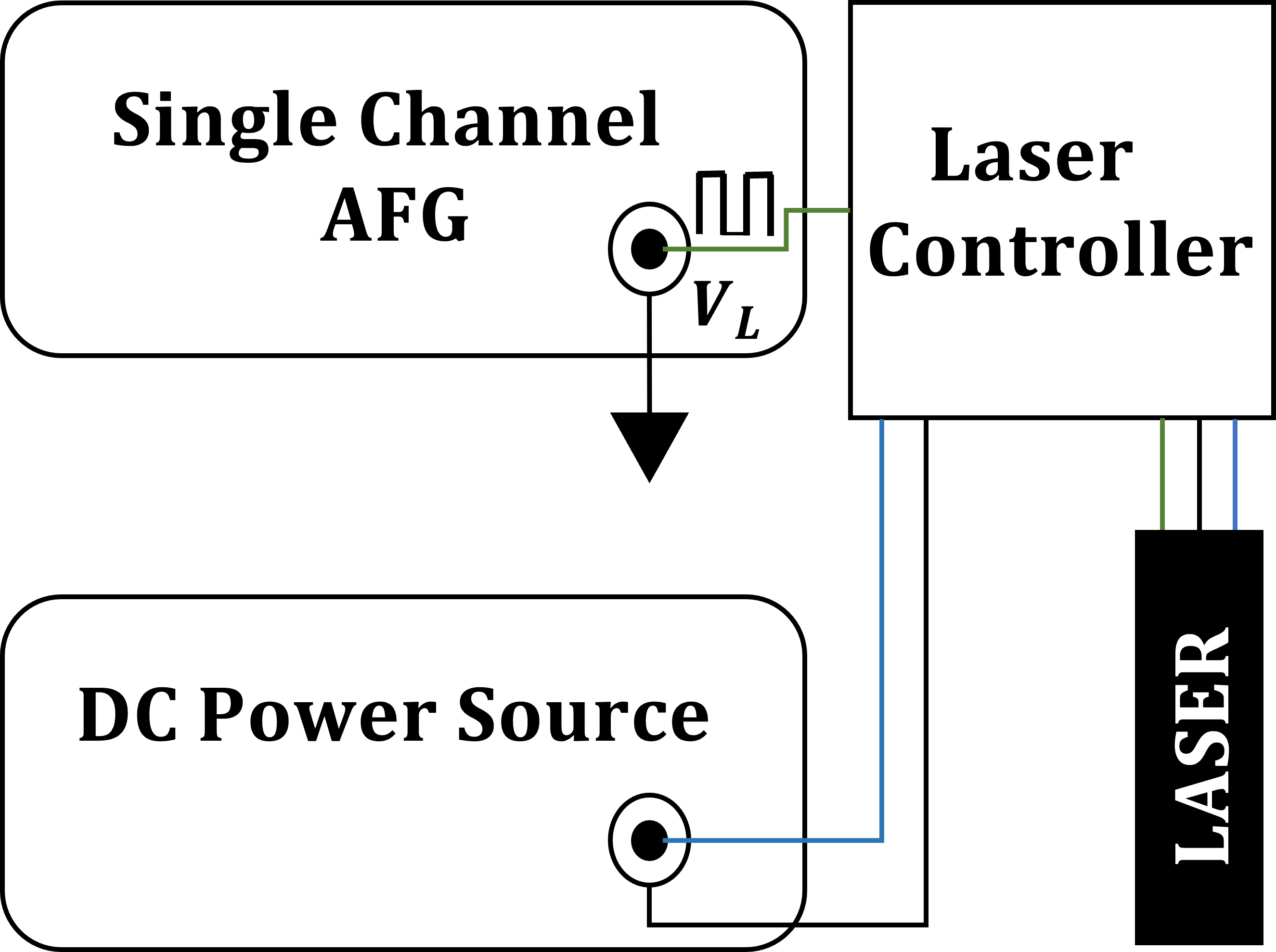}
      \caption{}
    \end{subfigure}\end{tabular}     
\caption{\justifying{(a) Experimental setup with the laser focused on the tip of the cantilever ($C_{\text{DUT}}$), (b) schematic of the experimental setup, and (c) a schematic of the laser setup.}}
\label{fig:experimental setup}
\end{figure*}

The capacitance bridge and the amplifier block were printed onto a copper PCB (Fig. \ref{fig:PCB and Cantilever photos}). See Appendix. \ref{appendix: Layout of the experiment - grounding} for details. In practice, the bridge voltage $V_{\text{Bridge}}(t)\sin(\omega_C t)$ was amplified using an inverting amplifier, giving the output voltage $V_{\text{Out}}(t)\sin(\omega_C t)$. $V_{\text{Out}}$ is the amplified signal amplitude monitored during the experiment. The time dependence in the amplitudes $V_{\text{Bridge}}$ and $V_{\text{Out}}$ are directly correlated to $\Delta C_{\text{DUT}}(t)$. High-quality equal-length (1m) coaxial cables were used to minimize stray capacitance and differences in phase delays in the signal paths. The inverting voltage amplifier built using an LF411C (Texas Instruments)\cite{LF411C} JFET input Op-Amp IC had a feedback ratio of $510\,\mathrm{k\Omega}/20\,\mathrm{k\Omega}$ and was powered using two 9V batteries.
$222.1 \,\mathrm{kHz}$ out-of-phase sinusoidal signals were sourced with amplitudes $V_{\text{DUT}}$ and $V_{\text{Ref}}$ using a two-channel AFG (Fig. \ref{fig:experimental setup}b). All signals were read out on a Digital Storage Oscilloscope with a $40\,\mathrm{GS/s}$ maximum sampling rate. The PCB was aligned on a horizontal x-y plane placed on a rubber pad for vibration isolation (Fig. \ref{fig:experimental setup}a). The setup was enclosed in a large cabinet, to prevent air currents and protect the eyes from scattered laser light. When the mean free path of air molecules becomes comparable to the cantilever separation, effects arising from the Knudsen force \cite{Knudsen} may affect measurements. Thus, it is essential to perform the experiment at ambient air pressure in the absence of air currents. 

We used a blue ($\lambda \simeq \,\mathrm{450\,nm}$) laser with an optical power $P_{\text{Avg}} = 0.4\pm0.02\,\mathrm{W}$, measured using a laser power meter when the laser was driven by a 50\% duty cycle digital signal. This type of laser is commonly used in laser engraving machines and is popular with DIY enthusiasts. The laser was controlled by a current drive circuit using a two-level digital pulse input (voltage $V_L$ switching between $0 \,\mathrm{V}$ and $5 \,\mathrm{V}$) with variable duty cycle and frequency $f_L$ (Fig. \ref{fig:experimental setup}c). The drive frequency for this model could be varied from DC up to $2 \,\mathrm{kHz}$, giving us a broad range of frequencies at which the cantilever could be driven. This square wave digital pulse with variable frequency and duty cycle was generated using a single channel AFG (Fig. \ref{fig:experimental setup}c). The laser was mounted on a clamp with adjustable z-alignment, and the height was adjusted such that the beam had the smallest spot size ($\leq 0.25\,\mathrm{mm}^2$) when focused on the tip of the cantilever (Fig. \ref{fig:experimental setup}a). The halo seen in the image (Fig. \ref{fig:experimental setup}a) is a diffraction ring caused by the focusing lens of the laser. Given the high contrast between the focused spot compared to the diffraction ring and the diffused light around the cantilever, it is reasonable to neglect the intensity of the latter.
\subsection{Amplifier gain and capacitance values}

\begin{figure}[h!]
\nolinenumbers
\includegraphics[width=0.9\linewidth]{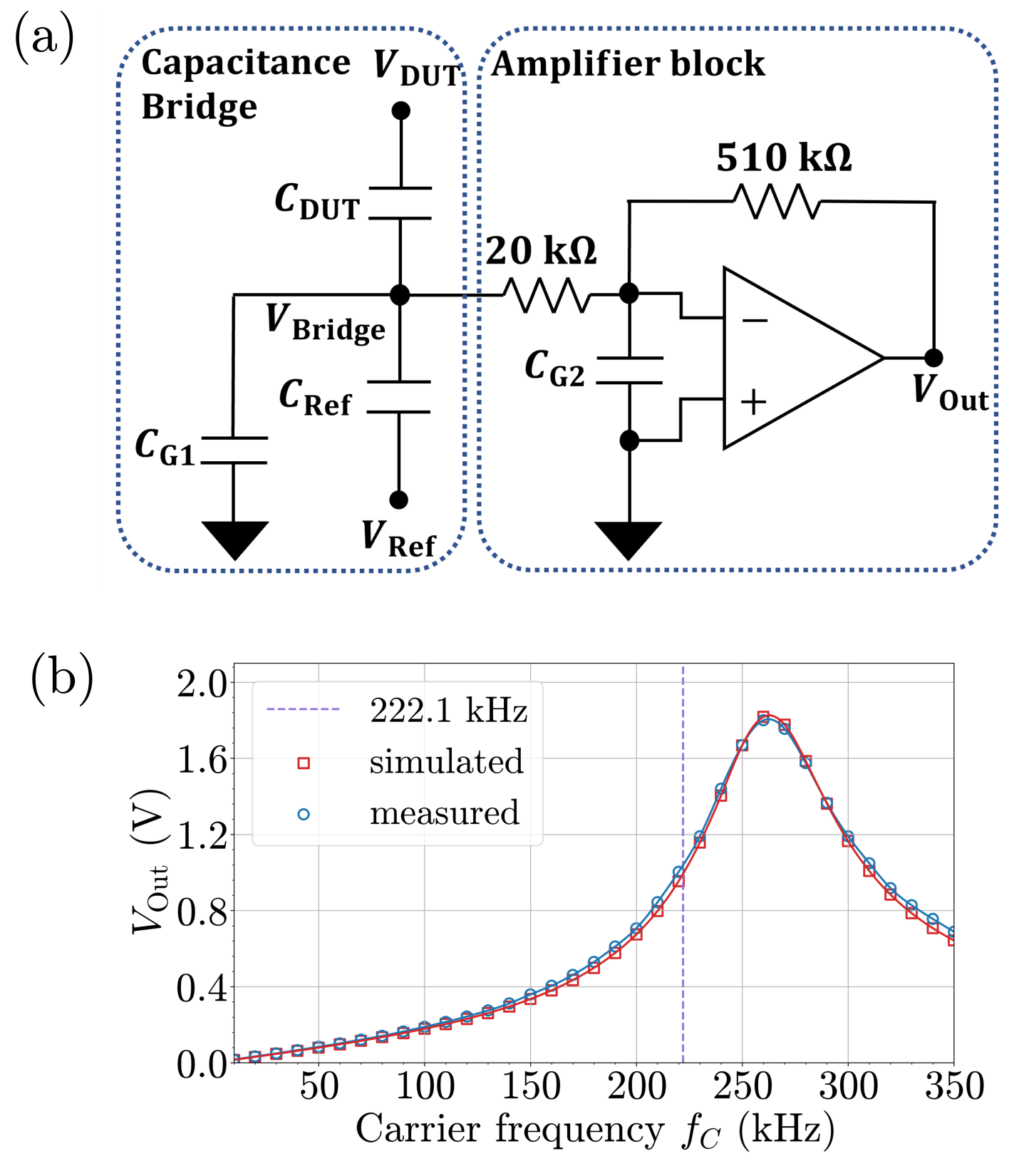}
\caption{\justifying{(a) Circuit schematic consisting the capacitance bridge and the amplifier block. We used LTSpice to simulate the circuit and match it with experiments to find capacitance $C_\text{G2}$. (b) Frequency dependence of amplification: measured ($\circ$) and simulated ($\square$) values of output voltage amplitude $V_{\text{Out}}$, for constant in-phase input signals with amplitudes $V_{\text{Ref}} = V_{\text{DUT}} = 1\,\mathrm{V_{pp}}$.}}
\label{fig: gain_matching}
\end{figure}

The electronic circuit can be divided into two blocks, namely, the capacitance bridge and the amplifier block (Fig. \ref{fig: gain_matching}a). In the absence of laser driving, the steady-state capacitances $C_{\text{Ref}}$, $C_{\text{DUT}}$, and $C_{\text{G1}}$ for the bridge circuit were measured to be $0.49\pm 0.01 \,\mathrm{pF}$, $0.47\pm 0.01 \,\mathrm{pF}$, and $8.2\pm 0.1 \,\mathrm{pF}$ using a capacitance meter. Due to the capacitive nature of the input to the inverting amplifier, the amplifier's gain varied strongly with the carrier frequency ($f_C$), showing a peak at $\simeq 260\,\mathrm{kHz}$ (Fig. \ref{fig: gain_matching}b). This could be understood as the convolution of the high-pass filter effect with the bridge capacitors and the gain bandwidth curve of the amplifier. The profile of the gain curve was measured experimentally by applying in-phase sinusoidal signals with amplitudes $V_{\text{DUT}} = V_{\text{Ref}} = 1\,\mathrm{V_{pp}}$ (peak to peak) and measuring $V_{\text{Out}}$ at frequencies from $10\,\text{kHz}$ up to $350\,\text{kHz}$ (Fig. \ref{fig: gain_matching}b). Near this gain peak, the Op-Amp output signal phase undergoes a $180^{\circ}$ shift, and operating around this point may lead to sustained oscillations of $V_{\text{Out}}$ and other instabilities. Thus, we chose $f_C = 222.1\,\mathrm{kHz}$ as the operating frequency where, $V_{\text{Out}} = 1\,\mathrm{V}\, (2\,\mathrm{V_{pp}})$.

We used LTSpice\cite{LTSpice} simulations to understand and verify experimentally measured results (see supplementary material for more details). We simulated the gain curve by performing a frequency sweep in LTSpice, using in-phase sinusoidal signals with amplitudes $V_{\text{DUT}}=V_{\text{Ref}}=1\,\mathrm{V_{pp}}$. Matching simulated and measured results (Fig. \ref{fig: gain_matching}b), the value of the effective lumped parasitic capacitance, $C_\text{G2}$ (see Fig. \ref{fig: gain_matching}a), seen at the input terminals of the Op-Amp was determined to be $6\,\text{pF}$. The value found is similar in magnitude to the value of the parasitic capacitance $C_\text{G1}$ which we measured.
\label{subsection: capacitance values}
\\
To simulate an oscillating cantilever capacitance in LTSpice, we used a voltage-controlled switch so that $C_{\text{DUT}}$ was modulated by $\Delta C_{\text{DUT}}$ at the frequency of a square wave source, which mimics the laser drive. See supplementary material for further details. The value of $\Delta C_{\text{DUT}}$ in the simulation was chosen to match experimental results discussed in the next section.
%--------------------------
\subsection{Steady state calibration with no laser}
\begin{figure}[ht]
\nolinenumbers
    \centering
    \includegraphics[width=\linewidth]{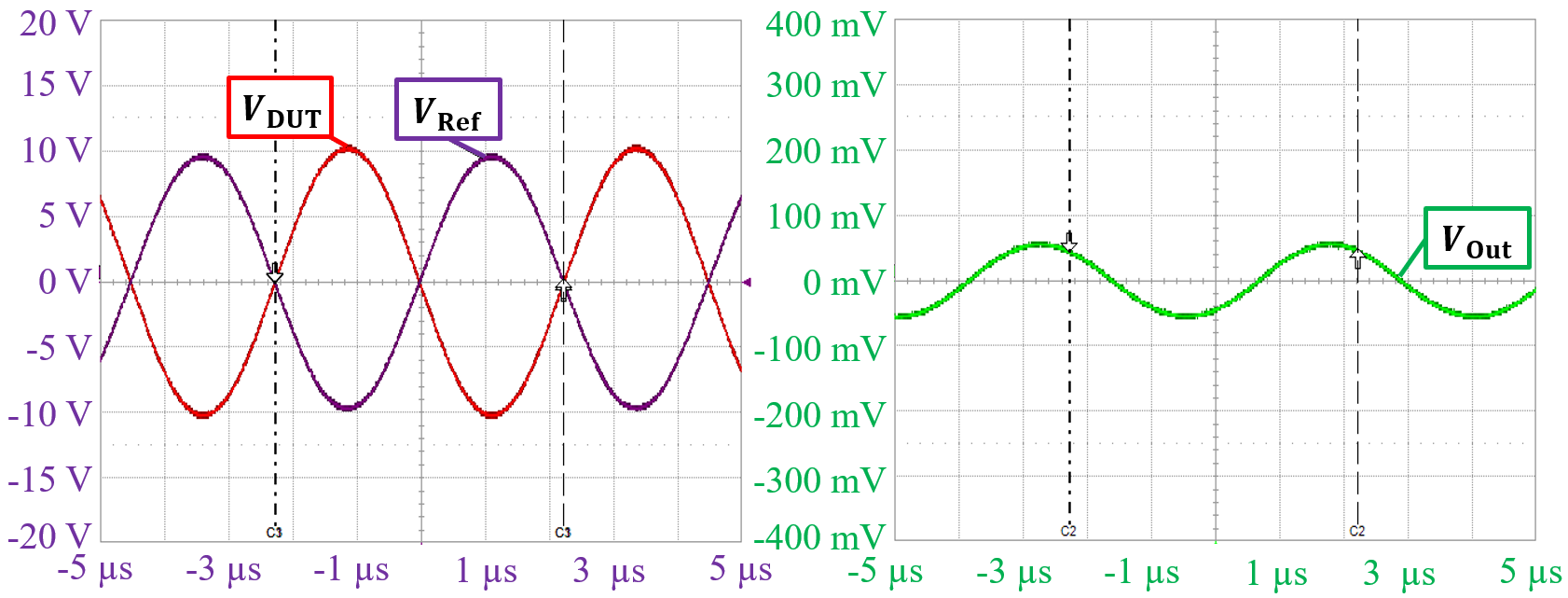}
    \caption{\justifying{Oscilloscope data for the balanced bridge: $V_{\text{Ref}}$ = $19\,\mathrm{V_{pp}}$ (left panel), $V_{\text{DUT}}$ = $20\,\mathrm{V_{pp}}$ (left panel), and $V_{\text{Out}}$ = $120\,\mathrm{mV_{pp}}$ (right panel).}}
    \label{fig:bridge balance expt}
\end{figure}
The first step while setting up the experiment is to balance the capacitance bridge (Fig. \ref{fig:bridge balance expt}). This was done by applying out-of-phase sinusoidal voltages at frequency $f_C = 222.1\,\mathrm{kHz}$ to the two input ports with $V_{\text{DUT}}$ = $20\,\mathrm{V_{pp}}$. The amplitude $V_{\text{Ref}}$, and relative phase $\eta$ were then varied until the amplitude $V_{\text{Out}}$ was minimized. Ideally, for exact destructive interference, ($\eta =\pi$), $V_{\text{Bridge}}$ should be zero. A small residual phase difference due to the minimum $1^\circ$ least count of the AFG causes a residual $V_{\text{Out}}$ (Fig. $\ref{fig:bridge balance expt}$b). In our experiment, $V_{\text{Ref}}$ = $19\,\mathrm{V_{pp}}$ balances the bridge ($V_{\text{Out}}\simeq 120\,\mathrm{mV_{pp}}$). Thus, Eq. (\ref{eq: balance}) gives $C_{\text{DUT}}/C_{\text{Ref}} = V_{\text{Ref}}/V_{\text{DUT}} = 0.95$.
This is in agreement with the measured values of bridge capacitances, where $C_{\text{DUT}}/C_{\text{Ref}} = 0.95\pm0.04$. The carrier frequency $f_C$ and signal amplitudes $V_{\text{Ref}}$ and $V_{\text{DUT}}$ were then kept constant for all further measurements.
%--------------------------------------------------%
%--------------------------------------------------%
\section{Results and Analysis}
\subsection{Determination of the cantilever's resonant frequencies}
\label{sec: Determination of the cantilever's resonant frequencies}
\begin{figure}[h!]
\includegraphics[width=\linewidth]{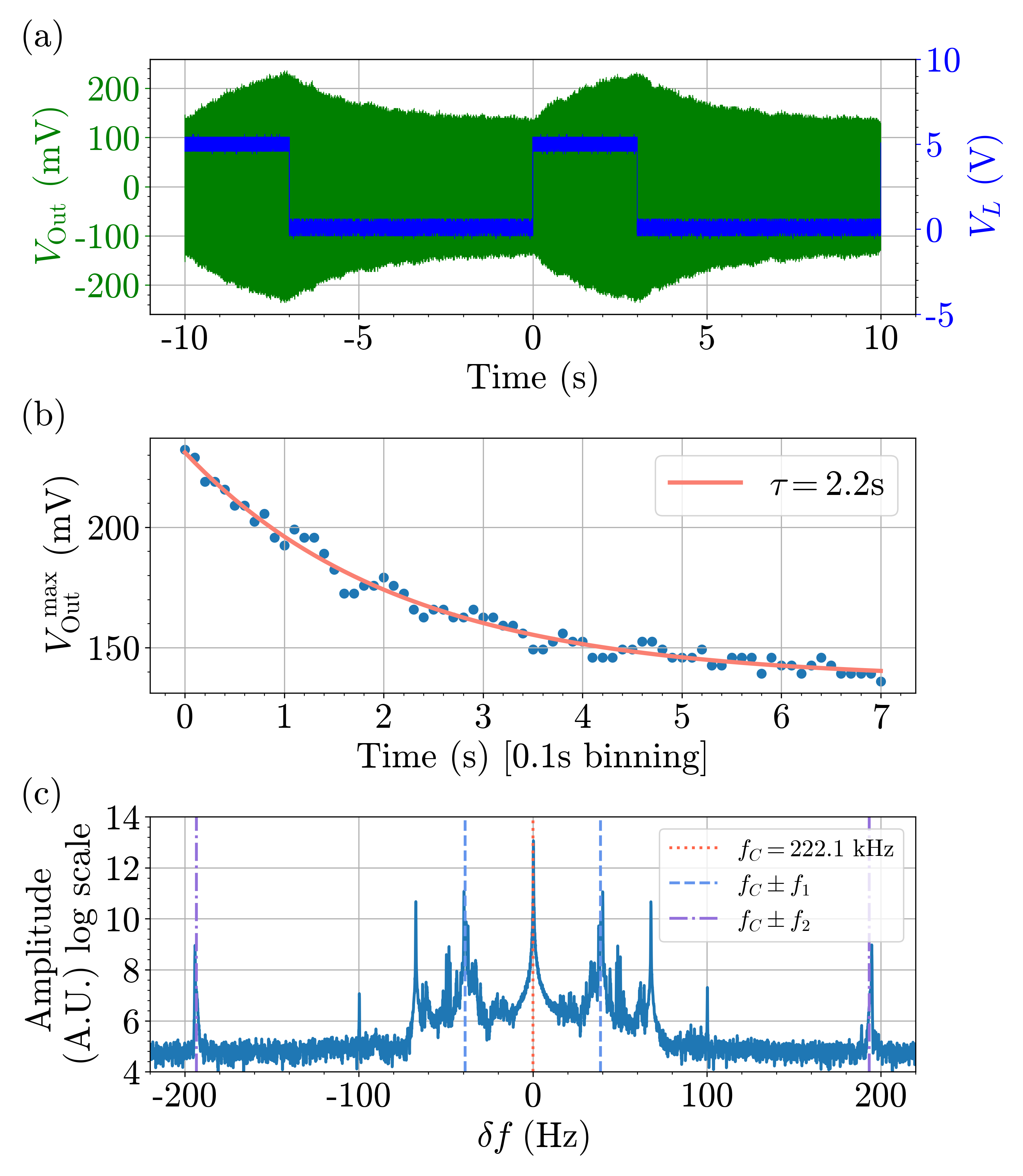}
\caption{\justifying{(a) Cantilever response for an off-resonance excitation at $f_L=0.1\,\mathrm{Hz}$: We measured $V_\text{Out}(t)$ when the laser was driven using a square wave voltage $V_L$ with duty cycle fo $30\%$. (b) Exponential fit to the envelope to extract the damping time constant $\tau$. (c) Discrete Fourier transform of $V_{\text{Out}}(t)$ plotted around the carrier frequency as a function of $\delta f = f-f_C$ showing the frequencies of resonant modes of the cantilever. For the Fourier transform, we use data of five periods spanning $50\,\mathrm{s}$.}}
\label{fig:off-resonance cantilever response}
\end{figure}

To characterize its dynamics, the cantilever was excited using a $f_L=0.1\,\mathrm{Hz}$ and 30\% duty cycle laser pulse (Fig. \ref{fig:off-resonance cantilever response}a). This measurement allows for the estimation of the damping constant and resonant frequencies. Since $f_L$ was kept much lower than the cantilever's expected first harmonic frequency $f_1^{\text{theory}}$, the cantilever could be thought of as being forced to displace below its mean position using a constant external force for $3\,\mathrm{s}$ and then left to oscillate freely back to equilibrium for $7\,\mathrm{s}$.\footnote{For this measurement, the exact value of $f_L$ and the duty cycle of the laser drive are not critical. To get a reliable exponential fit, it is important that the cantilever has enough time to relax back (close) to equilibrium when the laser is OFF, which is expected for a wait time longer than $\sim 3\tau \,(\text{reaches}\,95\%)$.} 
The envelope of $V_\text{Out}$ gives the cantilever's damping time constant $\tau = 2.2\,\mathrm{s} = 1/\Gamma$ as shown in to Fig. \ref{fig:off-resonance cantilever response}b (Eq. (\ref{eq: a7 decay})).
The discrete Fourier transform of the measured data, computed using a standard python FFT package (Fig. \ref{fig:off-resonance cantilever response}b) showed clear peaking not only at the carrier frequency $f_C = 222.1\,\text{kHz}$ but also $\simeq f_C \pm 39\,\mathrm{Hz}$ and $\simeq f_C \pm 193\,\mathrm{Hz}$. These values of $39\,\mathrm{Hz}$ and $193\,\mathrm{Hz}$ correspond to the first two flexural harmonics of the cantilever. We also observed peaks at $f_C\pm 50\,\mathrm{Hz}$ and $f_C\pm 100\,\mathrm{Hz}$ arising from the line frequency noise, as well as at $f_C\pm 70\,\mathrm{Hz}$, which we attribute to residual noise arising from differential coupling of the two halves of the bridge to ambient vibrational noise from air-conditioning compressors. This $70\,\mathrm{Hz}$ peak is also present in subsequent measurements with roughly the same relative magnitude, uncorrelated with the cantilever excitation.

\subsection{Resonant drive to measure radiation force}
\begin{figure}[h!]
\nolinenumbers
    \centering
    \includegraphics[width=\linewidth]{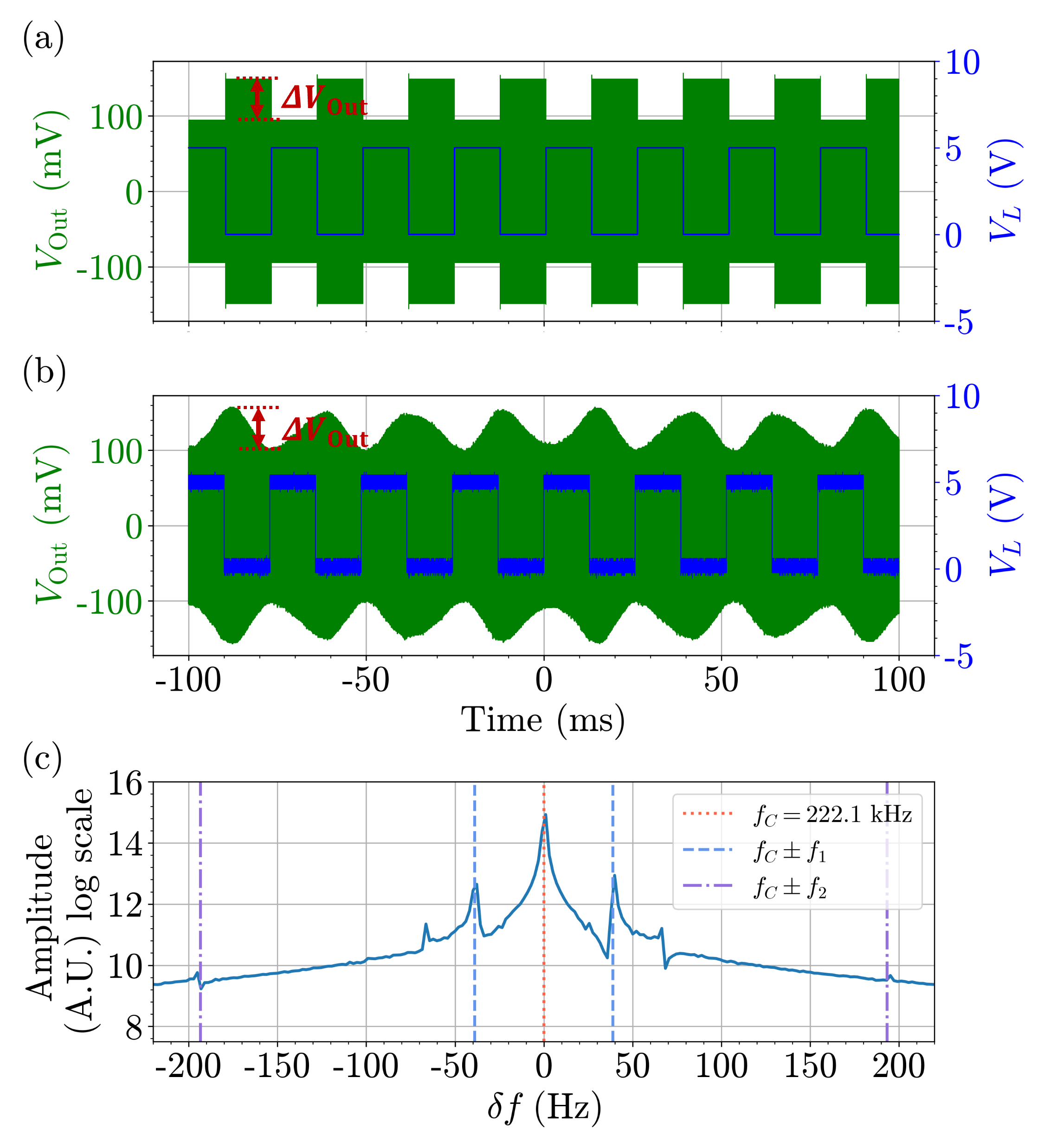}
    \caption{\justifying{Cantilever response for a resonant excitation at $f_L = f_1 = 38.881 \,\text{Hz}:$ (a) LTSpice simulation of the $V_{\text{Out}}$ with a $C_{\text{DUT}}$ varying by $2.6\,\mathrm{fF}$ when $V_L$ oscillates between $0\,\mathrm{V}$ and $5\,\mathrm{V}$. (b) Time-domain oscilloscope data, recording $V_\text{Out}$ in response to the laser driven with $V_L$. (c) Discrete Fourier transform of $V_{\text{Out}}(t)$ plotted around the carrier frequency as a function of $\delta f = f-f_C$. For the Fourier transform, we use data taken over 19 oscillation periods spanning $\sim 500\,\mathrm{ms}$.}}
\label{fig:resonant response at f1}
\end{figure}

Next, using a resonant drive with a duty cycle of $50\%$, we measured oscillations in the output amplitude $V_\text{Out}$. The cantilever's fundamental frequency was more accurately determined by fine-tuning $f_L$ around $39\,\mathrm{Hz}$ until the maximum amplitude was observed for the output signal. With our setup, we determined $38.881\,\mathrm{Hz}$, in agreement with the theoretical estimate (Eq. (\ref{eq: omega_1})).
Resonant oscillations were observed at this drive frequency (Fig. \ref{fig:resonant response at f1}b). The discrete Fourier transform of $V_{\text{Out}}$ showed prominent peaks at $f_C\pm f_{1}$ (dashed lines in Fig. \ref{fig:resonant response at f1}c).
% \clearpage
% \newpage

When excited at resonance, the maximum change in cantilever capacitance $\Delta C_\text{DUT}$ produces a corresponding change of $\Delta V_{\text{Out}}$ (labeled in Fig. \ref{fig:resonant response at f1}a and Fig. \ref{fig:resonant response at f1}b) in the output voltage. We simulate a circuit analogous to Fig. \ref{fig: gain_matching}a, but with a voltage ($V_L$) dependent $C_\text{DUT}$ which switches between capacitance values differing by $\Delta C_\text{DUT}$, which we tuned to match experiment. $\Delta C_{\text{DUT}}$ was obtained by matching $\Delta V_\text{Out}$ in simulations (Fig. \ref{fig:resonant response at f1}a) and experimental data (Fig. \ref{fig:resonant response at f1}b). The $\Delta V_{\text{Out}}$ signal envelope variation $53\pm{5}\,\mathrm{mV}$ (Fig. \ref{fig:resonant response at f1}a) was reproduced in simulations when $\Delta C_{\text{DUT}} = 2.6\pm0.2\,\mathrm{fF}$ (see supplementary material for more details).

For a square wave pulsed laser, the force exerted by the laser oscillates between 0 (OFF) and $2F_{\text{Rad}}$ (ON). Then, from the Fourier amplitude $F_1$ of driving force at frequency $f_1 = 38.881\,\mathrm{Hz}$ (Appendix \ref{appendix:A Dynamics of a damped driven cantilever}), $F_{\text{Rad}}$ can be determined:
\begin{equation}
    F_{\text{Rad}} = \frac{\pi}{4}\frac{\Delta C_{\text{DUT}}}{C_{\text{DUT}}} \frac{\rho \mathcal{A}_{\text{cs}}L_C\Gamma\omega_1d_0}{2\mathcal{J}} = 1.47\pm0.53\,\mathrm{\text{nN}}.
\end{equation}
Here, $d_0=0.44\pm0.08 \,\mathrm{mm}$ is the mean steady-state separation between the two plates of the air capacitor measured in-situ, and $\mathcal{J}$ is a dimensionless number derived in Appendix \ref{appendix:A Dynamics of a damped driven cantilever}. 
For brass ($a+2r = 1.5\pm 0.3$), \cite{brass_reflectance} this is equivalent to an average optical power $P_{\text{Avg}}$ given by:
\begin{equation}
    P_{\text{Avg}} = \left(\frac{cF_{\text{Rad}}}{1.5} \right)= 0.3\pm0.16\,\mathrm{W}.
    \label{eq: power}
\end{equation}
This agrees with the directly measured value (see Section \ref{subsec: laser and PCB}). A large deviation in values for $a+2r$ is considered since these cannot be measured in situ and may depend on the properties of the brass. This is the main source of error in the back-calculated power from the radiation force in Eq. (\ref{eq: power}).\footnote{Theoretically, $a+2r\in [1,2]$ with the extreme values corresponding to perfect absorption and perfect specular reflection.  Reference \cite{brass_reflectance} lists values for different brass alloys for variable wavelengths. These values for $\lambda = 450\,\mathrm{nm}$ were used to estimate $a+2r$.} If one uses a non-monochromatic source, reflectance would depend on the exact wavelength composition of the laser light.

\begin{figure}[h!]
\nolinenumbers
    \centering
    \includegraphics[width=\linewidth]{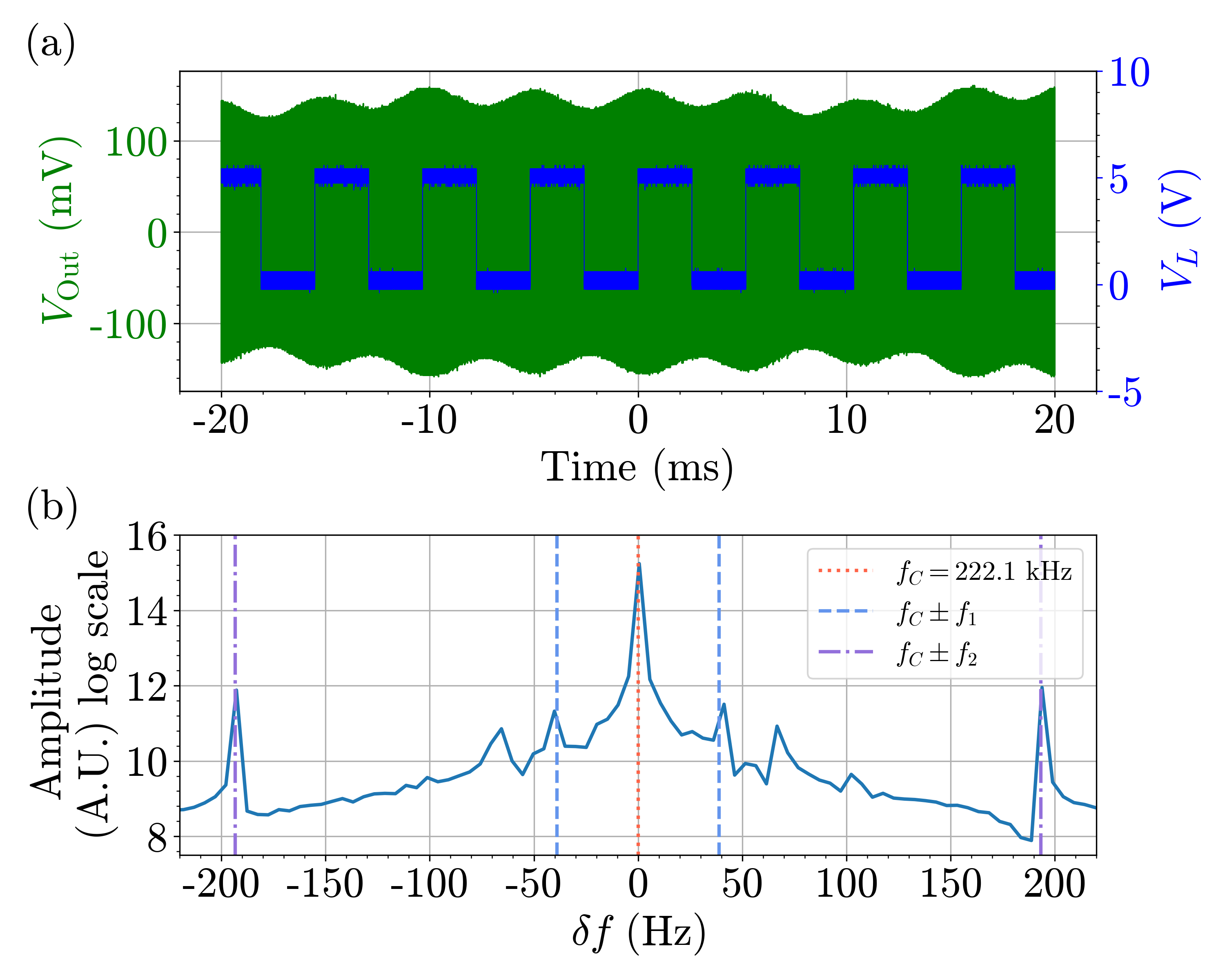}
    \caption{\justifying{Cantilever response for a resonant excitation at $f_L = f_2 = 193.361 \,\text{Hz}:$ (a) Time-domain oscilloscope data, recording $V_\text{Out}$ in response to the laser driven with $V_L$. (b)  Discrete Fourier transform of $V_{\text{Out}}(t)$ plotted around the carrier frequency as a function of $\delta f = f-f_C$. For the Fourier transform, we use data taken over 38 oscillation periods spanning $\sim 200\,\mathrm{ms}$.}}
\label{fig:resonant response at f2}
\end{figure}

We also observed resonant oscillations of the cantilever when excited at the second harmonic frequency $f_L = f_2 = 193.361\,\mathrm{Hz}$ (Fig. \ref{fig:resonant response at f2}). This value was experimentally determined with the procedure described above and deviates from the theoretical second harmonic frequency, which could be attributed to non-idealities due to heating and size non-uniformities that become significant for higher harmonics. \\

\section{Conclusion}
This work demonstrates unambiguous detection of mechanical vibrations of a thin metal cantilever caused by the radiation force, allowing detection of capacitance changes of the order of $\mathrm{fF}$ correlated to a few $\mathrm{nN}$ radiation forces. The experiment can be performed in ambient air conditions in a closed chamber, thus providing an accessible measurement of an otherwise hard-to-grasp theoretical aspect of Maxwell's equations.

The working principle of the interferometer used in this demonstration is the detection of extremely small capacitance changes, on the order of fF, using a capacitance-bridge configuration. In our case, these capacitance changes were induced by the radiation-force–driven motion of a cantilever. More generally, however, this setup functions as a very sensitive accelerometer. A multitude of interesting follow-up experiments can be designed. For instance, magnetic forces could be detected by measuring change in steady-state capacitance of a magnetic cantilever. Additionally, the experiment can be used as a way to perform spectroscopy of mechanical vibrational modes of more complex oscillators, so long as the mechanical motion can be coupled to capacitance changes. 

\section*{Supplementary Material}
Please click on this link to access the supplementary material, which includes the details about the LTSpice simulations used to model the circuit, and a 3D schematic of the layout of various instruments and electronic connections. Print readers can see the supplementary material at [DOI to be inserted by AIPP].

\section*{Author Contribution Statement and Data Availability}
The authors declare no conflict of interest. \textbf{D. Shah}: Formal Analysis (lead); Investigation (lead); Conceptualization (equal); Writing – original draft (lead); Writing – review and editing (equal). \textbf{P. Kumar}: Conceptualization (equal); Investigation (supporting). \textbf{P. Sarin}: Supervision (lead); Conceptualization (equal); Writing – original draft (supporting); Writing – review and editing (equal). \\

The data supporting the conclusions is available from the corresponding author upon reasonable request.
%-----------------------------------------------------------------------%
\appendix
\section{\\DYNAMICS OF A DAMPED DRIVEN CANTILEVER\cite{Repetto, LEISSA199183, Romaszko}}
\label{appendix:A Dynamics of a damped driven cantilever}
\begin{figure}[h!]
\nolinenumbers
    \centering
    \includegraphics[width=0.9\linewidth]{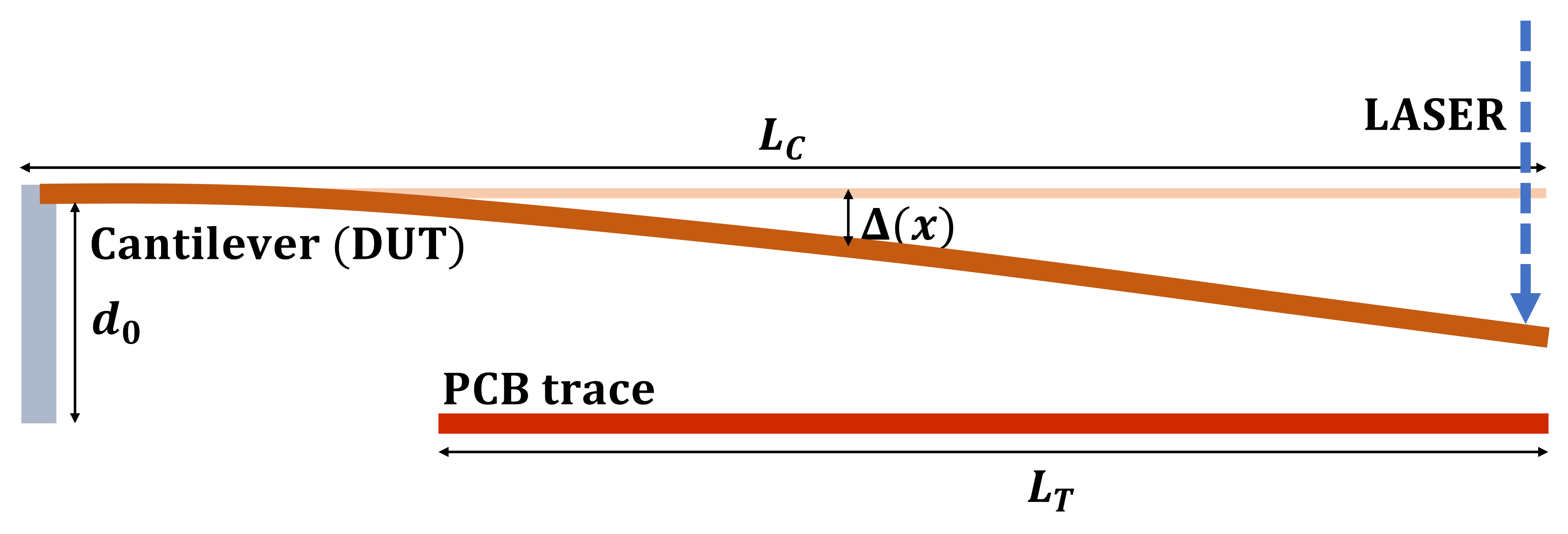}
    \caption{\justifying{Cantilever profile at maximum displacement form the mean position. The change in capacitance of the deflected cantilever is dominated by the first flexural eigenmode. This is the case when the beam is at its maximum displacement when pulsed at $f_1 = 38.881\mathrm{Hz}$.}}
    \label{fig:c_net calculation}
\end{figure}

In this section, we describe analytically, the dynamics of a cantilever fixed at one end. We consider a  cantilever with Young's modulus $Y$, cross-sectional area $\mathcal{A}_{\text{cs}} = bh$, and density $\rho$ (values listed in Table. \ref{table}). The cross section of the cantilever has width $b$, and thickness $h$. $M = b^3h/12$ is the area moment of inertia about the axis passing through the suspension point and along the width ($b$) of the cuboidal cantilever. The cantilever has a length $L_C$, part of which is in parallel to a copper PCB trace of length $L_T$, which forms the capacitor $C_{\text{DUT}}$ (Fig. \ref{fig:c_net calculation}).
\begin{table}[h!]
\centering
\begin{tabular}{p{3cm} c c}
\hline
\textbf{Quantity} & \textbf{Symbol} & \textbf{Value}\\
\hline
% Example entries:
Young' modulus & Y & $(1.0\pm 0.2)\cdot 10^{11}\,\mathrm{N/m^2}$  \\
Density & $\rho$ & $7575 \pm 19.9\,\mathrm{kg/m^2}$  \\
Cantilever width & $b$ & $1.0\pm 0.02 \,\mathrm{mm}$ \\
Cantilever thickness & $h$ & $50\pm 1 \,\mathrm{\mu m}$ \\
Cantilever length & $L_C$ & $2.740\pm 0.002 \,\mathrm{cm}$ \\
PCB Trace length & $L_T$ & $1.900\pm 0.002 \,\mathrm{cm}$ \\
\hline
\end{tabular}
\caption{\justifying{Cantilever dimensions and physical properties.}}
\label{table}
\end{table}

\textbf{No external drive:}
The deflection of the cantilever from its mean position, $\Psi(x,t)$, is given in the absence of an external drive and damping, by:
\begin{equation}
    \left(\frac{YM}{\rho \mathcal{A}_{\text{cs}}}\right)\frac{\partial^4\Psi}{\partial x^4}+\frac{\partial^2 \Psi}{\partial t^2} = 0.
    \label{eq: a1}
\end{equation}
Separation of variables $\Psi(x,t) = \chi(x)\xi(t)$ can be used to rewrite Eq. (\ref{eq: a1}) as:
\begin{equation}
    \left(\frac{YM}{\rho \mathcal{A}_{\text{cs}}\chi}\right)\frac{\mathrm{d}^4\chi}{\mathrm{d} x^4} = -\frac{1}{\xi}\frac{\mathrm{d}^2 \xi}{\mathrm{d} t^2} = \omega^2,
    \label{eq: variable separation}
\end{equation}
where $\omega^2$ is a constant. The general solution to the above equation can be written as a superposition of its free vibration eigenmodes $\phi_n(x)$:
\begin{equation}
\label{Eq: eigenmodes}
    \Psi(x,t) = \sum_{n=1}^\infty \phi_n(x)(A_n \sin(\omega_n t)+B_n\cos(\omega_n t)).
\end{equation}
Applying the boundary condition - one end fixed - the flexural eigenmodes can be shown to be:
\begin{equation}
\begin{split}
\label{Eq:eignemmodes_appendix}
    \phi_{n}(x)=& \cos(\beta_n x)-\cosh(\beta_n x)\\&-\kappa_n\left(\sin(\beta_n x)-\sinh(\beta_n x)\right),
\end{split}
\end{equation}
where, $\beta_n^{\;4}=\rho \mathcal{A}_{\text{cs}} \omega_n^{\;2}/YM$, $\kappa_n = (\cos(\beta_n L_C)+\cosh(\beta_n L_C))/(\sin(\beta_n L_C)+\sinh(\beta_n L_C))$, and $\beta_1L_C=1.875$, $\beta_2L_C=4.694$, $\beta_3L_C=7.855$. $\phi_n(0) = 0$, $\phi_n(L_C) = \pm 2$. The eigenmodes $\phi_n$ are orthogonal, with:
\begin{equation}
    \int^{L_C}_0\phi_n\phi_m \mathrm{d}x = L_C\delta_{nm}.
    \label{eq: orthonormality}
\end{equation}
In the presence of damping, the equation of motion is:
\begin{equation}
    \left(\frac{YM}{\rho \mathcal{A}_{\text{cs}}}\right)\frac{\partial^4\Psi}{\partial x^4}+\frac{\partial^2 \Psi}{\partial t^2} +\gamma\frac{\partial\Psi}{\partial t}= 0.
\end{equation}
Using a similar procedure as before, $\chi$ is spanned by $\phi_n(x)$ and for $\xi(t) = e^{i\alpha_n t}$, we have:
\begin{equation}
    \alpha_n = i\frac{\gamma}{2} \pm \omega_n\sqrt{1+\frac{\gamma^2}{4\omega_n^2}}.
    \label{eq: a7 decay}
\end{equation}
Thus, for a cantilever deflected from its rest position and left to oscillate, the oscillations decay exponentially as $e^{-\gamma t/2}$, which was measured experimentally. We call this decay rate $\Gamma = \gamma/2 = 1/\tau$ in Sec. \ref{sec: Determination of the cantilever's resonant frequencies}.

\textbf{Sinusoidal drive - radiation force:}
In the presence of an external sinusoidal force applied at one extremity the cantilever, the motion of the cantilever is governed by:
\begin{equation}
\label{eqn: damped driven oscillator}
     \left(\frac{YM}{\rho \mathcal{A}_{\text{cs}}}\right)\frac{\partial^4\Psi}{\partial x^4}+\frac{\partial^2 \Psi}{\partial t^2} +\gamma\frac{\partial\Psi}{\partial t}= \frac{F}{\rho \mathcal{A}_{\text{cs}}}\delta(x-L_C)e^{i\Omega t}.
\end{equation}
Here, $\Omega$ is the drive frequency and $F$ the amplitude of the drive.
For an under-damped oscillator, in steady state, $\Psi(x,t) = \sum_n A_n\phi_n(x)e^{i\Omega t}$, where $\phi_n(x)$ are the dimensionless eigenmodes (Eq. (\ref{Eq:eignemmodes_appendix})) and $A_n$ the associated amplitudes with dimensions of length. Substituting this in Eq. (\ref{eqn: damped driven oscillator}), and using Eq. (\ref{eq: variable separation}):
\begin{equation}
    \sum_n A_n\left[\omega_n^2 - \Omega^2 + i\gamma\Omega \right]\phi_n(x) e^{i\Omega t} = \frac{F}{\rho \mathcal{A}_{\text{cs}}}\delta(x-L_C)e^{i\Omega t}.
    \label{eq: driven amplitudes}
\end{equation}
Multiplying Eq. (\ref{eq: driven amplitudes}) by $\phi_m(x)$, integrating over the length of the cantilever by using the orthogonality of eigenmodes (\ref{eq: orthonormality}), and using $|\phi_m(L_C)|=2$, gives the amplitudes: 
\begin{equation}
    |A_n|=\frac{2F}{\rho A_\mathrm{cs}L_C}\cdot \frac{1}{\sqrt{\gamma^2\Omega^2 + (\omega_n^2 - \Omega^2)^2}}.
    \label{eq: eigen amplitudes}
\end{equation}
When driven at resonance $\Omega = \omega_1$ with a forcing amplitude $F_1$:
\begin{equation}
\label{eq: mod A1}
    |A_1| = \frac{F_1}{\Gamma\omega_1\rho \mathcal{A}_{\text{cs}} L_C}.
\end{equation}
Then, the maximum deflection of the cantilever from its mean position is $\Delta(x) = |A_1|\phi_1(x)$. Higher off-resonant harmonic amplitudes are highly suppressed $|A_n|<<|A_1|, \,\forall n>1$ and can be ignored.\footnote{Higher off-resonant harmonics are strongly suppressed and can be safely dropped in the analysis. For example, for our setup, for $\gamma \simeq 1\,\mathrm{s^{-1}}$, $f_1\simeq 40\,\mathrm{Hz}$, and $f_2\simeq 200\,\mathrm{Hz}$, $|A_2|\simeq 0.001|A_1|$.} The capacitance of the air capacitor when the cantilever is displaced to one of the extremes is $C^{'}_{\text{DUT}} = C_{\text{DUT}}\pm \Delta C_{\text{DUT}}/2$.

For small deflections ($|\Delta(x)|<<d_0$), the maximum change in the capacitance is:
\begin{equation}
    \frac{\Delta C_{\text{DUT}}}{2} =C_{\text{DUT}}\int^{L_C}_{L_C-L_T} \frac{|\Delta(x)|}{d_0}\frac{dx}{L_C}.
\end{equation}
Here, $C_{\text{DUT}}$, $L_C$, $L_T$, and $d_0$ were measured experimentally (Table. \ref{table}). Integration is over the region of overlap between the metal beam and the PCB trace, with $x=0$ being the position of the point from which the metal beam is suspended. The factor of half on the LHS is because experimentally, we determine the peak-to-peak value of the change in capacitance denoted as $\Delta C_{\text{DUT}}$. Using the expression for $\Delta(x)$ we get:
\begin{equation}
\label{eq: delta dut full expression}
    \frac{\Delta C_{\text{DUT}}}{2} = \frac{C_{\text{DUT}} |A_1|}{d_0}\int^{L_C}_{L_C-L_T} \frac{|\phi_1(x)|}{L_C}\mathrm{d}x.
\end{equation}
\begin{figure*}[ht]
    \centering
    \includegraphics[width=0.95\textwidth]{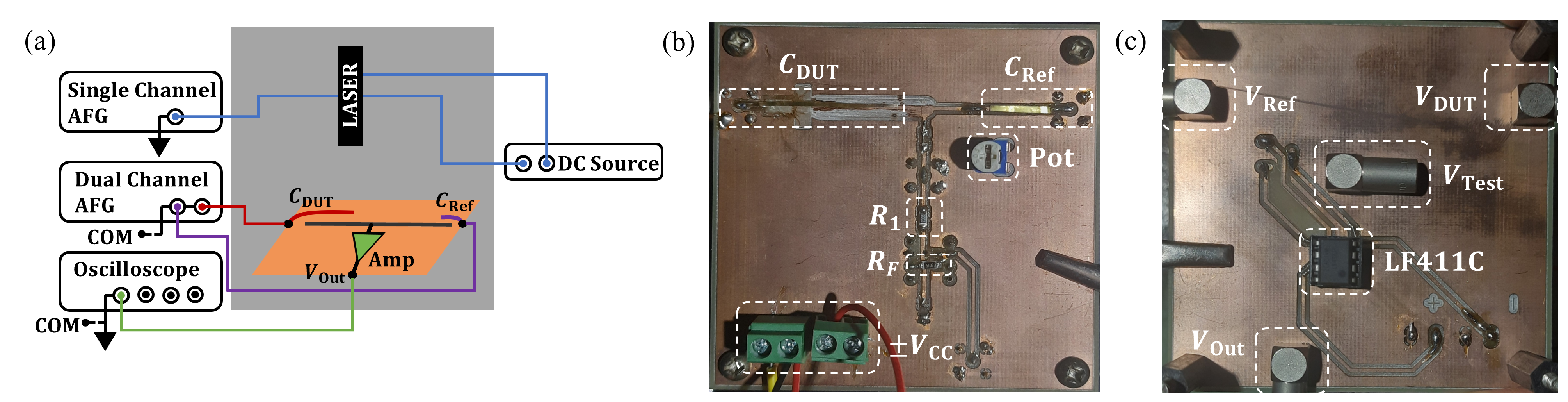} % Replace with your image
    \caption{\justifying{(a) Schematic of the experimental layout, with the PCB, the oscilloscope and the arbitrary function generator (AFG), and the laser setup with a DC power source. (b) Top side of the PCB with the two cantilever-based capacitors, a potentiometer (Pot) to adjust the offset of the Op-Amp, the resistors used for the Op-Amp feedback loop ($R_1$ and $R_F$), and the connectors for the $\pm 9V$ power for the Op-Amp ($\pm V_{\mathrm{CC}}$). (c) Bottom side of the PCB with the LF411C Op-Amp IC, the LEMO connectors for the three key ports to send and measure signals ($V_\text{DUT}, V_\text{Ref}, V_\text{Out}$), with a test port to check the input of the Op-Amp ($V_\text{Test}$). A double-sided PCB with points connected by vias allows for easy soldering.}}
    \label{fig:PCB_layout}
\end{figure*}
All quantities, except for $A_1$ (or equivalently $F_1$), in the above expression can be measured or obtained via simulation, allowing us to calculate the force amplitude $F_1$ (using \ref{eq: mod A1}):
\begin{equation}
\label{eq: F expression}
    F_1 = \left(\frac{\Delta C_{\text{DUT}}}{C_{\text{DUT}}}\right) \frac{\rho \mathcal{A}_{\text{cs}} L_C\Gamma\omega_1d_0}{2\mathcal{J}}.
\end{equation}
Here, $\mathcal{J}$ is the dimensionless integral in Eq. (\ref{eq: delta dut full expression}), which, in our case, computes to $0.75\pm0.06$, by substituting values for $L_C$, $L_T$, and the expression for $\phi_1(x)$. Further, the absolute maximum error in measurement can be estimated by accounting for relative errors in all quantities in Eq. (\ref{eq: F expression}).
% \begin{equation}
% \begin{split}
% \label{eq: error}
%     \frac{\delta F_1}{F_1} = &\frac{\delta \Delta C_{\text{DUT}}}{\Delta C_{\text{DUT}}}+\frac{\delta C_{\text{DUT}}}{C_{\text{DUT}}}+\frac{\delta \rho}{\rho}\\&+\frac{\delta \mathcal{A}_{\text{cs}}}{\mathcal{A}_{\text{cs}}}+\frac{\delta L_C}{L_C}+\frac{\delta d_0}{d_0} +\frac{\delta \mathcal{J}}{\mathcal{J}}.
% \end{split}
% \end{equation}
The relative errors in determining $\omega_1$ and $\Gamma$ are negligible compared to relative errors in measurement of $L_C$, $d_0$, $\mathcal{A}_{\text{cs}}$, and $\rho$ due to the resolution of the caliper (see Table. \ref{table}). The error in calculating $\mathcal{J}$ propagates from errors in $L_C$ and $L_T$.\\
A square digital light pulse exerts a force switching between 0 and $2F_{\text{Rad}}$, and an average force of $F_{\text{Rad}}$ for a $50\%$ duty cycle. For a pulsed laser at frequency $f_1=\omega_1/2\pi$, this force can be expanded in terms of Fourier components as:
\begin{equation}
    F(t) = F_{\text{Rad}}\left(1 + \frac{4}{\pi}\sin(\omega_1 t) + \frac{4}{3\pi}\sin(3\omega_1 t)+...\right).
\end{equation}
Thus, we experimentally measure $F_1=4F_{\text{Rad}}/\pi$.
%---------------------------------------------%
%---------------------------------------------%
\section{\\LAYOUT OF THE EXPERIMENT}
\label{appendix: Layout of the experiment - grounding}
Figure \ref{fig:PCB_layout} gives details of the experimental setup. Fig. \ref{fig:PCB_layout}a, shows the connections between electronic components. To minimize interference from the line frequency (50 Hz) on the power inputs, the grounding of the four instruments must be carefully designed. In practice, we operated the two AFGs with a floating power connection, that is, by supplying them power on a two-pin (live-neutral) connection instead of a standard  three-pin (live-neutral-earth) plug. One must check if the instruments are operable in this mode. Thus, the ground of the AFG driving the laser was kept separate from the ground of the AFG driving the sensing circuit. The DC power supply was a floating $12\,\mathrm{V}$ generated by a Keithley 2231A-30-3. The Digital Storage Oscilloscope we used was the only instrument that provided an absolute ground reference via a standard three-pin connector. The $0\,\mathrm{V}$ reference (COM in Fig. \ref{fig:PCB_layout}a) of the AFG driving the sensing circuit was connected to the ground through an implicit shared ground between the three cable connections on the PCB shown in Fig. \ref{fig:PCB_layout}c. The Op-Amp was powered by an independent pair of $9\mathrm{V}$ batteries connected in series to provide $+9\,\mathrm{V}$, $0\,\mathrm{V}$, and $-9\,\mathrm{V}$ (Fig. \ref{fig:PCB_layout}b). The Op-Amp's $0\,\mathrm{V}$ was also referred to the same global ground potential through the grounded copper surface of the PCB. Thus the integrity of a shared ground potential between $V_{\text{DUT}}$, $V_{\text{Ref}}$, and $V_{\text{Out}}$ was maintained. The global ground potential was kept separate from the laser-driving AFG, whose large TTL output ($5\,V_\mathrm{pp}$) at the drive frequency can inject spurious signals into the system through a ground loop. 

To allow for clean soldering, we used a two sided copper PCB with vias connecting the two sides. The dimensions of the PCB used were $70\,\text{mm}\times 70\,\text{mm}$. The front side (Fig. \ref{fig:PCB_layout}b) consisted the two bridge capacitances $C_\text{Ref}$ and $C_\text{DUT}$, the feedback resistors $R_1$ and $R_F$ of the amplifier, a potentiometer (Pot) for offset adjustment of the Op-Amp, and the power supply pins for the amplifier ($\pm V_\mathrm{CC}$). In addition, one sees a solder short connecting the capacitance bridge block to the amplifier block. This was designed to allow for testing the amplifier block using a test port independent of the capacitance bridge. On the back side, we soldered LEMO \cite{lemoconn} push-pull coaxial connectors for the input ($V_{\text{DUT}}$, $V_{\text{Ref}}$,), output ($V_{\text{Out}}$), and a test port ($V_{\text{Test}}$), because of their compact size, which keeps the PCB assembly small and easy to align. 

%\end{linenumbers}
%-----------------------------------------------------------------------%

\bibliographystyle{unsrt}
\bibliography{references}

\end{document}

\title{Supplemental information for\\ ``Measurement of electromagnetic radiation force using a capacitance-bridge interferometer"}

\author{Devashish Shah}
\affiliation{Department of Physics, Indian Institute of Technology Bombay, Mumbai, India}

\author{Pradum Kumar}
\affiliation{Department of Physics, Indian Institute of Technology Bombay, Mumbai, India}

\author{Pradeep Sarin}
\affiliation{Department of Physics, Indian Institute of Technology Bombay, Mumbai, India}
\date{\today}

\maketitle

\section{LTSpice simulations for the experiment}
\begin{figure}[h!]
    \centering
    \includegraphics[width=0.8\linewidth]{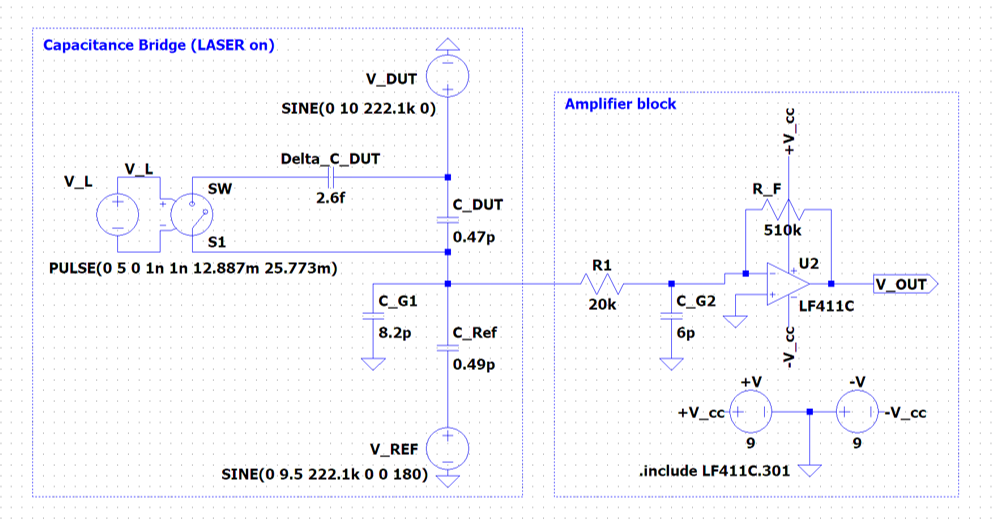}
    \caption{\justifying{LTSpice circuit for simulations consisting of the bridge circuit and the amplifier block.}}
    \label{fig:LTSpice circuit}
\end{figure}

We use LTSpice simulations to model the electronic circuit being used for the experiment. The circuit consists of two main parts, the capacitance bridge and the amplifier block (Fig. \ref{fig:LTSpice circuit}). 

\begin{table}[h!]
\centering
\begin{tabular}{p{2.5cm} l}
\hline
Label & Component description\\
\hline
% Example entries:
$V_\mathrm{Ref}$ &  AC voltage source with amplitude $V_{\mathrm{DUT}}$ - sinusoidal, frequency $222.1\,\mathrm{kHz}$, and phase $0^\circ$\\
$V_\mathrm{DUT}$& AC voltage source with amplitude $V_{\mathrm{Ref}}$ - sinusoidal, frequency $222.1\,\mathrm{kHz}$, and phase $180^\circ$\\
$V_\mathrm{L}$& $5\,\mathrm{V_{pp}}$ AC voltage source - square wave, 50\% duty cycle, and frequency $38.881\,\mathrm{Hz}$\\
$C_\mathrm{Ref}$&$0.49\,\mathrm{pF}$ capacitor - the reference air capacitor\\
$C_\mathrm{DUT}$ &$0.47\,\mathrm{pF}$ capacitor - the device under test (DUT) air capacitor \\
$C_\mathrm{G1}$&$8.2\,\mathrm{pF}$ capacitor - lumped parasitic capacitance to ground \\
SW & An ideal switch - it is open when V\_TTL is $0\,\mathrm{V}$ it and is a perfect short when V\_TTL is $5\,\mathrm{V}$\\
$\Delta C_\mathrm{DUT}$& $2.6\,\mathrm{fF}$ capacitor - models variable $C_\mathrm{DUT}$ realized in the experiment\\
\hline
\end{tabular}
\caption{\justifying{Capacitance bridge: component name and description.}}
\label{tableS1}
\end{table}

\begin{table}[h!]
\centering
\begin{tabular}{p{2.5cm} l}
\hline
Label & Component description\\
\hline
% Example entries:
$R_1$ & $20\,\mathrm{k\Omega}$ resistor - resistor connecting the bridge to the input ($-$ pin) of the Op-Amp\\
$R_{F}$& $510\,\mathrm{k\Omega}$ resistor - feedback resistor connecting the Op-Amp output to the Op-Amp input ($-$ pin)\\
$C_\mathrm{G2}$&$6\,\mathrm{pF}$ capacitor - lumped parasitic between the Op-Amp input pins ($+$ pin is grounded)\\
LF411C & Spice Op-Amp model for Texas Instruments LF411C IC \cite{LF411C}\\
+V\_cc & $+9\,\mathrm{V}$ DC voltage source - powers the Op-Amp\\
-V\_cc & $-9\,\mathrm{V}$ DC voltage source - powers the Op-Amp\\
\hline
\end{tabular}
\caption{\justifying{Amplifier block: component name and description.}}
\label{tableS2}
\end{table}

\subsection{Matching the amplification curve to get $C_\mathrm{G2}$}
Capacitances $C_\mathrm{DUT}$, $C_\mathrm{Ref}$, and $C_\mathrm{G1}$ were measured in-situ using a capacitance meter. However, $C_\mathrm{G2}$, the lumped parasitic capacitance between the two Op-Amp input pins is difficult to measure reliably. To estimate its value, we performed a frequency sweep for the circuit shown in Fig. \ref{fig:LTSpice circuit}. At this stage, we disconnect the $\Delta C_\mathrm{DUT}$, to mimic the steady state condition of the bridge circuit, and apply, equal ($V_\mathrm{DUT} = V_\mathrm{Ref}= 1\,\mathrm{V_{pp}}$), in phase sinusoidal voltages to the two bridge capacitors. We then plot the $V_\mathrm{Out}$ as a function of frequency for different values of $C_\mathrm{G2}$, and find the best agreement with the experiment for $C_\mathrm{G2} = 6\,\mathrm{pF}$ (Fig. 4b).  

\subsection{Simulating laser drive using a variable $\Delta C_\mathrm{DUT}$}
Once $C_\mathrm{G2}$ is determined, the simulation circuit can be used to determine the change in DUT capacitance induced by oscillations driven by the laser on resonance. For this, we use the circuit topology shown in Fig. \ref{fig:LTSpice circuit}, with $V_\mathrm{Ref} = 19\,\mathrm{V_{pp}}$ and $V_\mathrm{DUT} = 20\,\mathrm{V_{pp}}$, and perform time domain analysis over a $500\,\mathrm{ms}$ window. Our measurements showed a maximum change in amplitude $V_\mathrm{Out}$ of $53\pm 5\,\mathrm{mV}$. The two maximum and minimum values of $V_\mathrm{Out}$ observed correspond to the extreme cantilever deflections from the mean position. When the cantilever bends towards the PCB trace, the capacitance reaches a maximum of $C'_\mathrm{DUT}=C_\mathrm{DUT}+\Delta C_\mathrm{DUT}/2$ (this is the case shown in Fig. 9). For the opposite extreme, the cantilever bends away from the PCB trace, where $C'_\mathrm{DUT}=C_\mathrm{DUT}-\Delta C_\mathrm{DUT}/2$. Thus, the change in the amplitude $V_\mathrm{Out}$ corresponds to the full peak-to-peak change $\Delta C_\mathrm{DUT}$ of the device capacitance. We reproduce this in simulations by choosing $\Delta C_\mathrm{DUT}$ around $2.55\,\mathrm{fF}$ (Fig. 7a).

\section{Further details regarding the experimental setup}
\begin{figure}[h!]
    \centering
    \includegraphics[width=\linewidth]{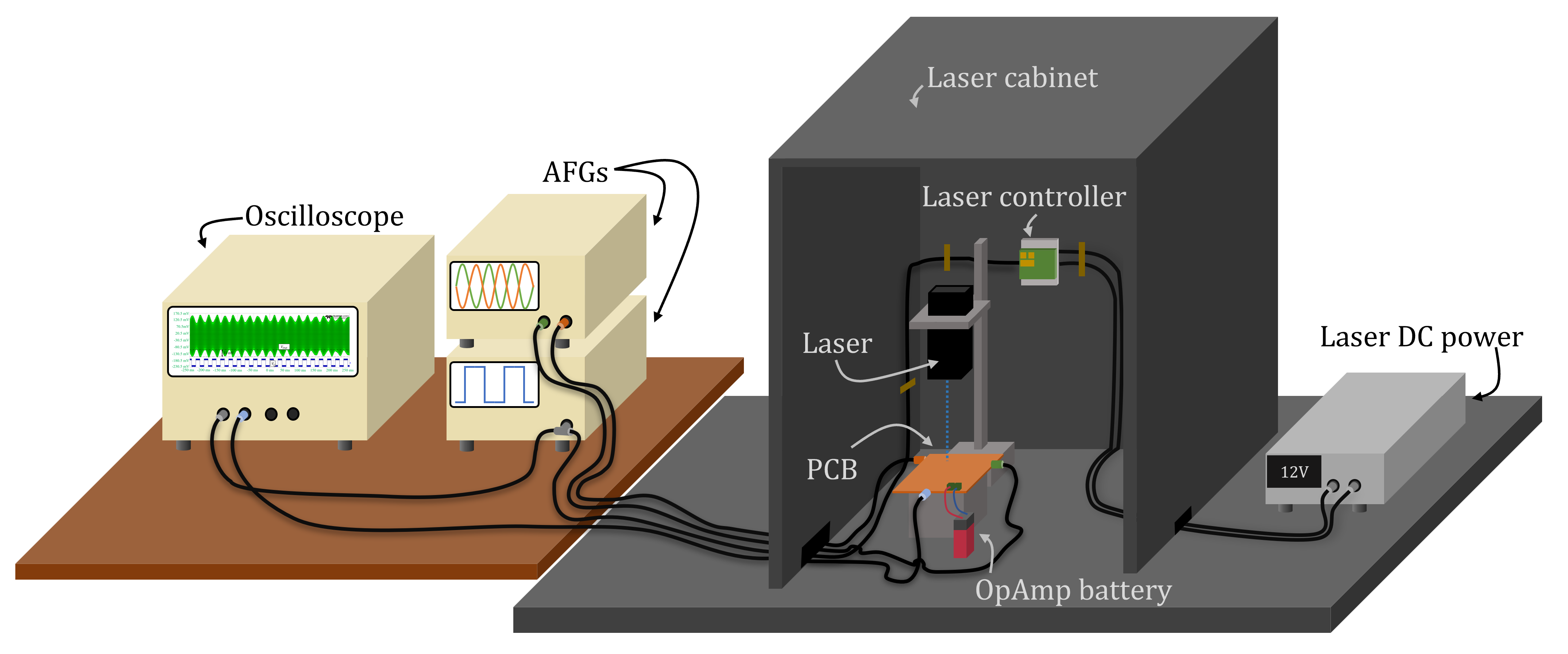}
    \caption{\justifying{Layout of the experiment.}}
    \label{fig:3D layout}
\end{figure}

Figure \ref{fig:3D layout} shows the layout of various instruments and the connections in detail. The laser and PCB are housed in a cabinet with opaque Aluminum walls (painted black), which along with the power source for the laser is placed on an optical bench. The oscilloscope and arbitrary function generators are placed on another table, with coaxial cables connecting them to the PCB and laser, through holes in the casing. 

\bibliography{references}